\documentclass[apjl]{emulateapj}
\shortauthors{Kaltenegger, Selsis}
\shorttitle{Biomarkers set in context}

\begin{document}

\title{Biomarkers set in context}

\author{L. Kaltenegger}
\affil{Harvard-Smithsonian Center for Astrophysics,
60 Garden St, Cambridge, MA 02138, USA} 
\email{lkaltenegger@cfa.harvard.edu}
\and
\author{F. Selsis}
\affil{CRAL-ENS, 46 Allee d'Italie, 69364 Lyon, Cedex 7} 

\begin{abstract}
In a famous paper, Sagan et al. \cite{Sagan1993} analyzed a spectrum of the Earth taken by the Galileo probe, 
searching for signatures of life. They concluded that the large amount of $O_2$ and the simultaneous presence 
of $CH_4$ traces are strongly suggestive of biology. The detection of a widespread red-absorbing pigment with 
no likely mineral origin supports the hypothesis of biophotosynthesis. The search for signs of life on possibly 
very different planets implies that we need to gather as much information as possible in order to understand 
how the observed atmosphere physically and chemically works. 

The Earth-Sun intensity ratio is about $10^{-7}$ in the thermal infrared ($\sim$ 10 $\mu$m), and about $10^{-10}$ in the 
visible ($\sim$ 0.5 $\mu$m). The interferometric systems suggested for Darwin and the Terrestrial Planet Finder 
Interferometer (TPF-I) mission operates in the mid-IR (5 - 20 $\mu$m), the coronagraph suggested for Terrestrial 
Planet Finder Coronagraph (TPF-C) in the visible (0.5 - 1 $\mu$m). For the former it is thus the thermal emission 
emanating from the planet that is detected and analyzed while for the later the reflected stellar flux is 
measured. The spectrum of the planet can contain signatures of atmospheric species that are important for 
habitability, like $CO_2$ and $H_2O$, or result from biological activity ($O_2$, $O_3$, $CH_4$, and $N_2O$). Both spectral 
regions contain atmospheric bio-indicators: $CO_2$, $H_2O$, $O_3$, $CH_4$, and $N_2O$ in the thermal infrared, and $H_2O$, 
$O_3$, $O_2$, $CH_4$ and $CO_2$ in the visible to near-infrared. The presence or absence of these spectral features 
will indicate similarities or differences with the atmospheres of terrestrial planets and are discussed in 
detail and set into context with the physical characteristics of a planet in this chapter \footnote{Extrasolar Planets: Formation, Detection and Dynamics, Eds R.Dvorak, Copyright 2007 WILEY-VCH, ISBN: 987-3-527-40671-5}.
\end{abstract}

\keywords{Biomarkers, Extrasolar planets, Habitabiity, Search for Life}

\section{Introduction}
\label{Biomarkers}
Over 260 giant exoplanets have already been detected, and hundreds, perhaps thousands more, are anticipated in the coming years.
The detection and characterization of these exoplanets will begin to fill in a gap in the astrophysical description of the 
universe; the chain of events between the first stages of star formation and the evident mature planetary systems. The nature 
of these planets, including their orbits, masses, sizes, constituents, and likelihood that life could develop on them, can be 
probed by a combination of observations and modeling. So far, our main detection methods from the ground, Radial Velocity and Transit 
Search, are biased towards big planets orbiting close to their parent star because they are easier to detect. Those planets 
produce a bigger, more frequent signal than a small planet further away from its parent star. Even so, the number of smaller Extrasolar Giant Planets (EGP)
found, indicates a trend towards smaller masses.

The present stage of exoplanet observations can be characterized as one in which information is being gathered principally by 
indirect means, whereby the photons that we measure are from the star itself, or a background star, or a mixture of the star and 
planet. Indirect techniques include radial velocity, micro-lensing, transits, and astrometry. These indirect observations are of
great value, giving us measures of the planet mass, orbital elements, and (for transits) the sizes as well as indications of the
constituents of the extreme upper atmospheres like the detection of sodium in the upper atmosphere of HD209458b \cite{Charbonneau2002}. 
Recent detections of super Earths by Udry et al.\cite{Udry2007}, Rivera et al. \cite{Rivera2005} and Beaulieu et al. \cite{Beaulieu2006} imply that Earth-mass planets might be 
common \cite{Gould2006}. Current and future space missions like CoRoT (CNES, \cite{Rouan1998}) and Kepler (NASA \cite{Borucki1997}), 
will give us statistics on the number, size, period and orbital distance of planets, extending to terrestrial planets on the 
lower mass range end. 

In the next stage of exoplanet observations, we may hope to have direct observations, in which most of the measured photons 
are reflected or emitted by the planet itself. A fundamental part of the problem of directly detecting the planet with its 
feeble light in the glare of the strong parental stellar flux, is the huge contrast (see Fridlund, this volume). Direct 
techniques include coronagraphic imaging at visible wavelengths, and interferometric imaging in the thermal infrared. With 
direct photons in the visible and thermal infrared wavelength band, and depending on the Signal to Noise Ratio, we can 
characterize a planet in terms of its size, albedo, its atmospheric gas constituents, total atmospheric column density, clouds, 
surface properties, land and ocean areas and general habitability. As discussed in Traub et al. \cite{Traub2006}, full characterization 
requires the synergy of both direct and indirect measurements. Direct detection of photons from giant exoplanets can be 
implemented using current space based telescopes like HST and Spitzer. Such studies have led to the detection of infrared 
emission from several transiting hot Jupiters (see e.g. \cite{Charbonneau2005}; \cite{Deming2005}, \cite{Harrington2006}) where 
the planetary signal is the difference between the flux from a star plus planet versus the flux from the star alone. Photons 
from Earth-like planets in the habitable zone (HZ) (see \cite{Kasting1993}) around their parent star are beyond the 
capabilities of these telescopes and require future missions like Darwin and Terrestrial Planet Finder (TPF).

In this chapter we discuss the biomarkers at different wavelengths and the potential of each signatures. 
In section 4.2 to section 4.8 we focus on what makes a habitable planet using Earth as our example and discuss surface, clouds and 
biosignature evolution over geological time. Section 4.9 concentrates on planets around different stars, section 4.10 
focuses on abiotic sources of potential biosignatures and section 4.11 and section 4.12 show how to interpret biosignatures by setting 
measurements in context with physical characteristics of a planet. Section 4.13 summarizes the chapter.

\section{Biomarkers}\index{Biomarkers}
{\it Biomarkers} (or biosignature) is used here to mean detectable species, or set of species, whose presence at significant 
abundance strongly suggests a biological origin \cite{DesMarais2002}. This is for instance the case for the couple $CH_4$ + $O_2$.
{\it Bio-indicators} are indicative of biological processes but can also be produced abiotically in significant quantities. 
Our search for signs of life is based on the assumption that extraterrestrial life shares fundamental characteristics with life on Earth, 
in that it requires liquid water as a solvent and has a carbon-based chemistry \cite{Owen1980}; \cite{DesMarais2002}. 
Life on the base of a different chemistry is not considered here because of the vast possible signatures unknown life-forms produce in 
their atmosphere that are so far unknown. Therefore we assume that extraterrestrial life is similar to life on Earth in its use 
of the same input and output gases, that it exists out of thermodynamic equilibrium, and that it has analogs to bacteria, plants, 
and animals on Earth \cite{Lovelock1975}. 

The first step of exploration of terrestrial extrasolar planets will be a space mission that can detect and record low resolution 
spectra (see Fig. 1) of extrasolar planets like Darwin and Terrestrial Planet Finder (TPF). $O_2$, $O_3$, $CH_4$ are good biomarker candidates 
that can be detected by a low-resolution (Resolution $<$ 50) spectrograph. There are good biogeochemical and thermodynamic reasons
for believing that these gases should be ubiquitous byproducts of carbon-based biochemistry, even if the details of alien 
biochemistry are significantly different than the biochemistry on Earth. Note that life can also exist without producing either 
$O_2$ or $CH_4$. Even so, we need to understand the abiotic sources of biomarkers better, so that we can identify when it might 
constitute a false positive for life detection, when abiotic sources could produce high quantities of a species we understand 
as a biomarker on Earth. The theoretical modeling research goals are to explore the plausible range of habitable planets and to 
improve our understanding of the detectable ways in which life modifies a planet on a global scale.

\section{Biomarker signatures in different wavelength ranges}\index{Biomarker signatures in different wavelength ranges}

With arbitrarily high signal-to-noise (SNR) and spatial and spectral resolution, it is relatively straightforward to remotely 
ascertain that Earth is a habitable planet, replete with oceans, a greenhouse atmosphere, global geochemical cycles, and life.  
The interpretation of observations of other planets with limited signal-to-noise ratio and spectral resolution as well as absolutely 
no spatial resolution, as envisioned for the first generation missions like TPF-C and Darwin/TPF-I, will be far more challenging.

To search for signs of life with low resolution and limited information implies that we need to gather as much information as 
possible in order to understand what we will see. The following step by step approach can be taken to set the system in context.
After detection, we will focus on main properties of the planetary system, its orbital elements as well as the presence of an 
atmosphere using the light curve of the planet or/and a crude estimate of the planetary nature using very low-resolution information 
(3 or 4 channels)\cite{Traub2002}. Then a higher resolution spectra will be used to identify the compounds of the planetary atmosphere, establish 
the temperature and radius of the observed exoplanet. In that context, we can then test if we have an abiotic explanation of all
compounds seen in the atmosphere of such a planet. If we do not, we can work with the exciting biotic hypothesis. Second 
generation space mission will then investigate those targets in more detail to improve our understanding of those environments. 
The results of a first generation mission will most likely consist in an amazing scope of diverse planets that will set planet 
formation, evolution as well as our planet in an overall context.

The thermal infrared concepts, the Darwin and the Terrestrial Planet Finder Interferometer (TPF-I), and the visible wavelength 
concepts, the Terrestrial Planet Finder Coronagraph (TPF-C), are designed to detect terrestrial exoplanets, and to measure the 
color and spectra of terrestrial planets, giant planets, and zodiacal dust disks around nearby stars (see, e.g., \cite{Leger1993}, 
\cite{Beichman1999}; \cite{Beichman2006}; \cite{Fridlund2000}; \cite{Kaltenegger2004}; \cite{Kaltenegger2005}; 
\cite{Borde2006}). These missions have the explicit purpose of detecting other Earth-like worlds, analyzing their 
characteristics, determining the composition of their atmospheres, investigating their capability to sustain life as we know it,
and searching for signs of life. They are a first step to characterize a vast number of unknown fascinating planetary worlds. 
Their respective resolution is envisioned to about 25 for Darwin and TPF-I and about 70 for TPF-C. These missions also have the 
capacity to investigate the physical properties and composition of a broader diversity of planets, to understand the formation 
of planets and interpret potential biosignatures. 

The range of characteristics of planets is likely to exceed our experience with the planets and satellites in our own Solar 
System. Earth-like planets orbiting stars of different spectral type might evolve differently \cite{Selsis2000}; 
\cite{Segura2003}, \cite{Segura2005}. Models of such planets need to consider the changing atmosphere structure, as well as 
the interior structure of the planet (see e.g. \cite{Valencia2006}, \cite{Sotin2007}). One crucial factor in interpreting 
planetary spectra is the point in the evolution of the atmosphere when its biomarkers become detectable. Studies of individual 
constituent effects on the spectrum and resolution estimates were previously discussed for Earth in Des Marais et al. \cite{DesMarais2002}. 
These calculations were for a current atmospheric temperature structure, but different abundances of chemical species. 
Spectra of the Earth exploring temperature sensitivity (a hot house and cold scenario) and different singled out stages of its 
evolution (e.g., \cite{Schindler2000}, \cite{Selsis2000}, \cite{Pavlov2000}, \cite{Traub2002}, \cite{Segura2003}, 
\cite{Kaltenegger2006}) as well as the evolution of the expected spectra 
of Earth \cite{Kaltenegger2006} produce a variety of spectral fingerprints for our own planet. Those spectra will be used as 
part of a big grid to characterize any exoplanets found. This also influences the design requirements for a spectrometer 
\cite{Kaltenegger2006} to detect habitability.

%                                                One column figure
%----------------------------------------------------------- Fig1
\begin{figure}
\centering
\includegraphics[width=8.cm]{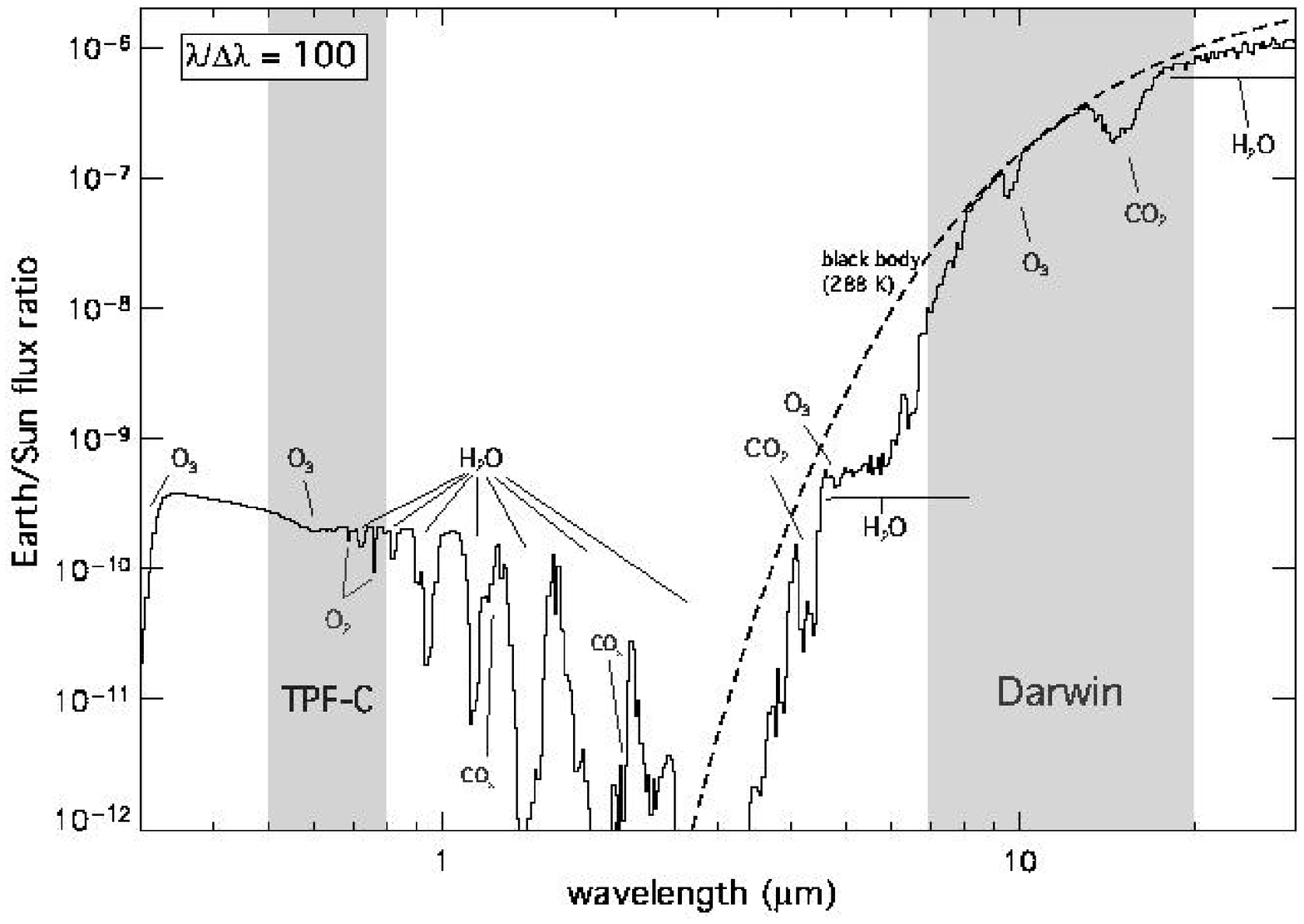}
\includegraphics[width=8.cm]{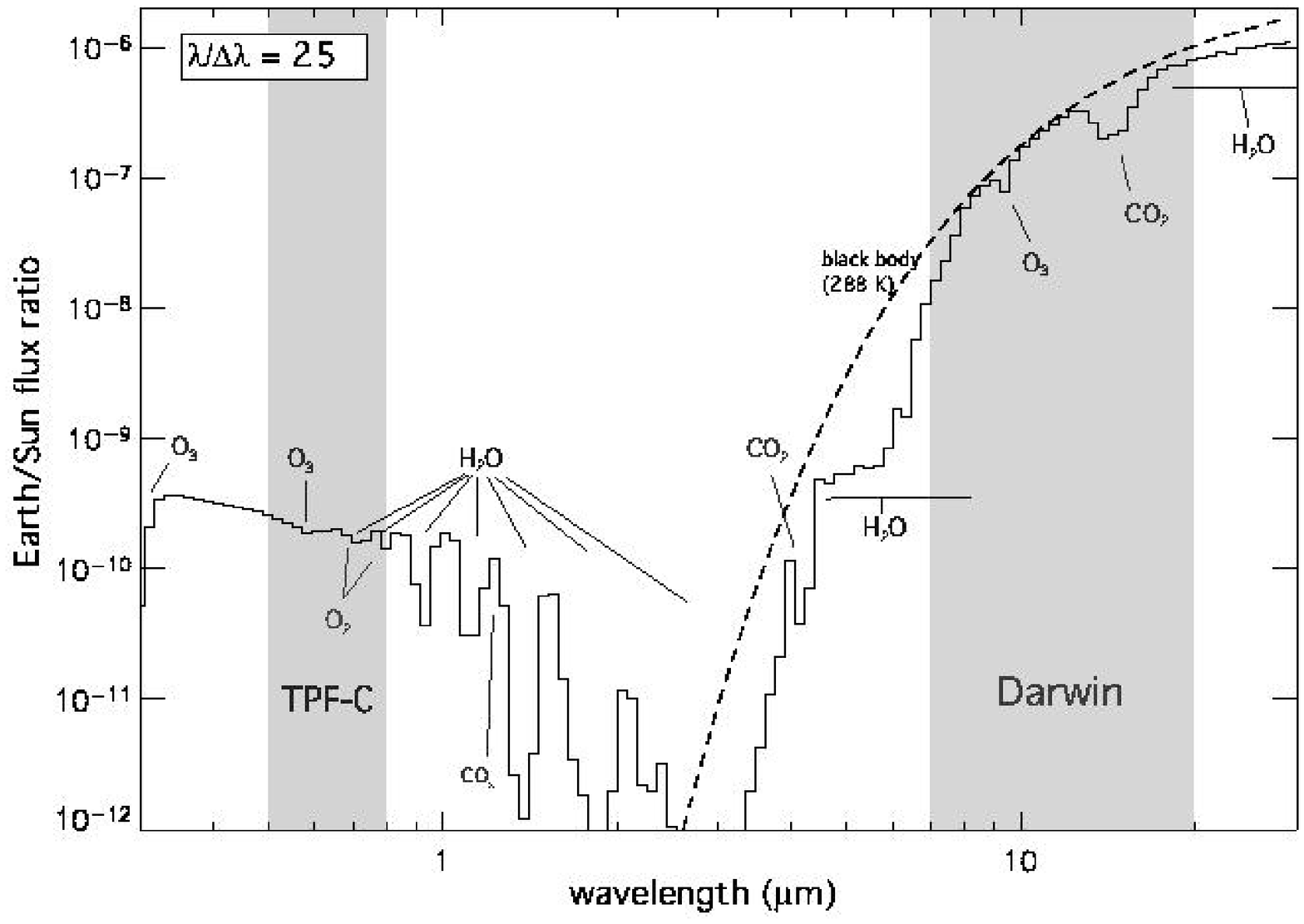}
\caption{Synthetic spectra of the Earth from UV to IR shown in two different resolutions (R=100 and 25) representing the 
proposed resolution for TPF-C (R=70) and Darwin TPF-I (R=25). The intensity is given as a fraction of solar intensity. 
The atmospheric features as well as the spectroscopic range of Darwin/TPF-I and TPF-C is indicated \cite{Paillet2006}.}
\end{figure}
%
%__________________________________________________________________

\section{Potential Biomarkers}
\index{Potential Biomarkers}
Oxygen in high abundance is a promising bio-indicators. Oxygenic photosynthesis, which by-product is molecular oxygen extracted 
from water, allows terrestrial plants and photosynthetic bacteria (cyanobacteria) to use abundant $H_2O$, instead of having to 
rely on scarce supplies of electron donor to reduce $CO_2$, like $H_2$ and $H_2S$. With oxygenic photosynthesis, the production of 
the biomass becomes limited only by nutriments and no longer by energy (light in this case) nor by the abundance of electron 
doners. Oxygenic photosynthesis at a planetary scale results in the storage of large amounts of radiative energy in chemical 
energy, in the form of organic matter. For this reason, oxygenic photosynthesis had a tremendous impact on biogeochemical cycles 
on Earth and eventually resulted in the global transformation of Earth environment.
Reduced gases and oxygen have to be produced concurrently to be detectable in the atmosphere, as they react rapidly with each 
other. Thus a detectable concentration of $O_2$ and/or $O_3$ and of a reduced gas like $CH_4$ can be considered as a signature of 
biological activity. The spectrum of the Earth have exhibited a strong infrared signature of ozone for more than 2 billion years,
and a strong visible signature of $O_2$ for a period of time between 2 and 0.8 billion years (depending on the 
required depth of the band for detection and also the actual evolution of the $O_2$ level). This difference is due to the fact 
that a saturated ozone band appears already at very low levels of $O_2$ ($10^{-4}$ ppm) while the oxygen line remains unsaturated at 
values below 1 Present Atmospheric Level (PAL). Note that the non-detection of $O_2$ or $O_3$ on an exoplanet cannot be interpreted as the absence of life. 

$N_2O$ is produced in abundance by life but only in trace amounts by natural processes. Nearly all of Earth's $N_2O$ is produced by 
the activities of anaerobic denitrifying bacteria. $N_2O$ would be hard to detect in Earth's atmosphere with low resolution, as 
its abundance is low at the surface (0.3 ppmv) and falls off rapidly in the stratosphere. On a low-$O_2$ early Earth, its 
abundance would be even smaller because it photolyzes rapidly in the near ultraviolet. As signs of life in themselves $H_2O$ and 
$CO_2$ are secondary in importance because although they are not indicators of its presence, they are raw materials for life and 
thus necessary for planetary habitability. 

In the visible reflected spectrum of earth, the detectable signatures of biological
activity in low resolution are water vapour and molecular oxygen (mainly the
0.76 $\mu$m band), in the near-IR $CO_2$ and $CH_4$ are detectable at concentrations
significantly higher than on current earth. 

In the mid-IR thermal emission spectra of earth the combined detection of the 9.6 $\mu$m 
$O_3$ band, the 15 $\mu$m $CO_2$ band and the 6.3 $\mu$m $H_2O$ band or its rotational band that extends from 12 $\mu$m out into the 
microwave region \cite{Selsis2002} indicating habitability. The 9.6 $\mu$m $O_3$ band is a very nonlinear
indicator of $O_2$: First, for the present atmosphere, low resolution spectra of this band show little change with 
the $O_3$ abundance because it is strongly saturated.  Second, the apparent depth of this band remains nearly constant as $O_2$ 
increases from 0.01 PAL of $O_2$ to 1 PAL \cite{Segura2003}. 
The primary reason for that is
that the strastopheric warming decreases with the abundance of ozone, making the $O_3$ band deeper. The depth of the saturated $O_3$
band is indeed determined by the temperature difference between the surface-clouds continuum and the ozone layer.  
$CH_4$ is not readily identified using low resolution spectroscopy for present-day Earth, but the methane feature at 7.66 $\mu$m in 
the IR is easily detectable at higher abundances (see e.g. 100x abundance, Fig.5 epoch 4), provided of course that the spectrum 
contains the whole band and that a high enough SNR. Taken together with molecular oxygen, abundant $CH_4$ can indicate biological 
processes (see also \cite{Lovelock1975}, \cite{Sagan1993}, \cite{Segura2003}). Depending on the degree of oxidation of a planet's 
crust and upper mantle non-biological mechanisms can also produce large amounts of $CH_4$ under certain circumstances. 

$N_2O$ is produced by life but only in negligible amounts by abiotic processes. There are potentially three weak $N_2O$ features in 
the thermal infrared at 7.75 $\mu$m and 8.52 $\mu$m, and 16.89 $\mu$m. Methane and nitrous oxide, have features nearly overlapping in 
the 7 $\mu$m region, and additionally both lie in the red wing of the 6 $\mu$m water band. Although its abundance is less than 1 ppm 
in Earth atmosphere, the 7.75 $\mu$m shows up in a medium resolution infrared spectrum. Spectral features of $N_2O$ would become 
more apparent in atmospheres with more $N_2O$ and/or less $H_2O$ vapor. On a low-$O_2$ early Earth, its abundance would be even smaller 
because it photolyzes rapidly in the near ultraviolet. Segura et al. (2003) have calculated the level of $N_2O$ for different $O_2$ 
levels and found that, although $N_2O$ is a reduced species, its levels decreases with $O_2$. This is due to the fact that a decrease in 
$O_2$ produces an increase of $H_2O$ photolysis resulting in the production of more hydroxyl radicals (OH) responsible for the 
destruction of $N_2O$.

There are other molecules that could, under some circumstances, act as excellent biomarkers, e.g., the manufactured chloro-fluorocarbons 
($CCl_2F_2$ and $CCL_3F$) in our current atmosphere in the thermal infrared waveband, but their abundances are too low to be spectroscopically 
observed at low resolution.

\section{A habitable Planet}
\index{A habitable Planet}

The circumstellar Habitable Zone (HZ) is defined as the region around a star within which starlight is sufficiently intense to 
maintain liquid water at the surface of the planet, without initiating runaway greenhouse conditions vaporizing the whole water 
reservoir and, as a second effect, inducing the photodissociation of water vapor and the loss of hydrogen to space (Fig.2), see J. Kasting \cite{Kasting1993},\cite{Kasting1997} 
for a detailed discussion. On an Earth-like planet where the carbonate-silicate cycle is at work, the level of $CO_2$ in the 
atmosphere depends on the orbital distance: $CO_2$ is a trace gas close to the inner edge of the HZ but a major compound in the 
outer part of the HZ. Earth-like planets close to the inner edge are expected to have a water-rich atmosphere or to have lost 
their water reservoir to space. As the HZ is defined for surface conditions only, chimio-lithotrophic life, which metabolism does 
not depend on the stellar light, can still exist outside the HZ, thriving in the interior of the planet where liquid water is 
available. Such metabolisms (at least the ones we know on Earth) do not produce $O_2$ and rely on very limited sources of energy 
(compared to stellar light) and electron donors (compared to $H_2O$ on Earth). They mainly catalyze reactions that would occur at 
a slower rate in purely abiotic conditions and they are thus not expected to modify a whole planetary environment in a detectable 
way. 

%----------------------------------------------------------- Fig2a
\begin{figure}
\centering
\includegraphics[width=8.4cm]{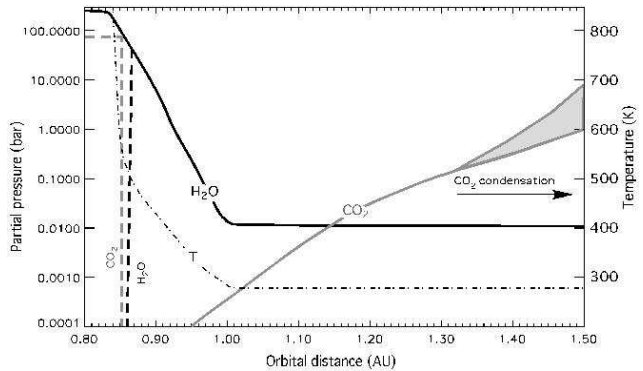}
\caption{The mean surface temperature (TS) and partial pressure of $CO_2$ and $H_2O$ as a function
of the orbital distance on a habitable planet within the habitable zone ((data adapted from \cite{Kasting1993} 
\cite{Forget1997} (partial pressure (left y-axis) and TS (right y-axis)).}
\end{figure}
%
%__________________________________________________________________

As we said, partial pressure of $CO_2$ and $H_2O$ at the surface of an Earth-like habitable planet is a function of the orbital 
distance a, within the HZ and for the present solar luminosity. We assume here that the planet contains 1 terrestrial ocean of 
superficial water and that carbonate-silicate cycle is at work, controlling the $CO_2$ level in equilibrium with a surface 
temperature at about 290 K, for a $>$ 1 AU. For a $>$ 1.3 AU, $CO_2$ condensates in the atmosphere producing $CO_2$ clouds that can 
affect the Temperature - $CO_2$ coupling significantly. For a $<$ 0.93 AU, $H_2O$ becomes a major atmospheric compound and is rapidly 
lost to space after UV photolysis. Thus, a Venus-like fate (no $H_2O$ remaining and a massive $CO_2$ build-up) is likely in this 
inner part of the HZ. This is one of the first theories we can test with a first generation space mission. However, the limits 
of the HZ are known qualitatively, more than quantitatively. This uncertainty is mainly due to the complex role of clouds but 
also three-dimensional climatic effects not yet included in the modeling. Thus, planets slightly outside the computed HZ 
could still be habitable, while planets at habitable orbital distance may not be habitable because of their size or chemical 
composition.

\section{Oxygen and ozone production on Earth}
Owen (1980) suggested searching for $O_2$ as a tracer of life. In the particular case of Earth, $O_2$ is fully produced by the 
biosphere. Less than 1 ppm comes from abiotic processes \cite{Walker1977}. Cyanobacteria and plants are responsible for this 
production by using the solar photons to extract hydrogen from water and using it to produce organic molecules from $CO_2$. 
This metabolism is called oxygenic photosynthesis. The reverse reaction, using $O_2$  to oxidize the organics produced by 
photosynthesis, can occur abiotically when organics are exposed to free oxygen, or biogically by eukaryotes breathing $O_2$ and 
consuming organics. Because of this balance, the net release of $O_2$ in the atmosphere is due to the burial of organics in 
sediments (see Fig.3). Each reduced carbon buried, releases a free $O_2$ molecule in the atmosphere. This net release rate is also balanced by 
weathering of fossilized carbon when exposed to the surface. The oxidation of reduced volcanic gasses such as $H_2$ and $H_2S$ also 
accounts for a significant fraction of the oxygen losses. The atmospheric oxygen is recycled through respiration and 
photosynthesis in less than 10 000 yrs. In the case of a total extinction of Earth biosphere, the atmospheric $O_2$ would 
disappear in a few million years.

Ozone is produced in the atmosphere by a unique chemical reaction: $ O + O_2 + M -> O_3 + M $, where $M$ is any compound. This reaction 
is not very efficient as it requires at the same time a high enough pressure (because it is a 3 bodies reaction), and oxygen 
atoms that are produced at lower pressures where photolysis of $O_2$ by UV can occur. Ozone can be efficiently destroyed by a 
large number of reactions, dominated, in the Earth's atmosphere, by catalytic cycles involving trace species such as hydrogenous 
compounds (H , OH , $HO_2$ ), nitrogen oxides ($NO_X$) and chlorine compounds ($ClO_X$). These species have various origins and their 
amount depends on the nature and the intensity of the bio-productivity, the thermal profile of the atmosphere, human pollution, 
and many other parameters. Without these compounds, an atmosphere made of $N_2$ and $O_2$ only would contain 10 times more $O_3$. 
The column density of $O_3$ in the atmosphere depends weakly on the abundance of $O_2$, the mean opacity of the 9.6 $\mu$m band 
remaining $>$ 1 for $O_2$ abundance as low as $10^{-3}$ PAL \cite{Leger1993}, \cite{Segura2003}.

The abundance of $O_3$ and its observability also vary with the spectral distribution of the incoming stellar radiation. For 
atmospheric compositions similar to the Earth's one, numerical simulations show that $O_3$ increases with the UV flux 
\cite{Selsis2000}, \cite{Segura2003}. The depth of the $O_3$ feature depends however on the difference beween the brightness 
temperature of the continuum (given by the temperature of the surface and/or the clouds) and the temperature of the ozone layer 
(where the opacity of $O_3$ is 1). Because of this complex coupling, less ozone means less warming and still a deep feature, while 
more ozone produces more warming and a shallow feature.

%----------------------------------------------------------- Fig2b
\begin{figure}
\centering
\includegraphics[width=8.4cm]{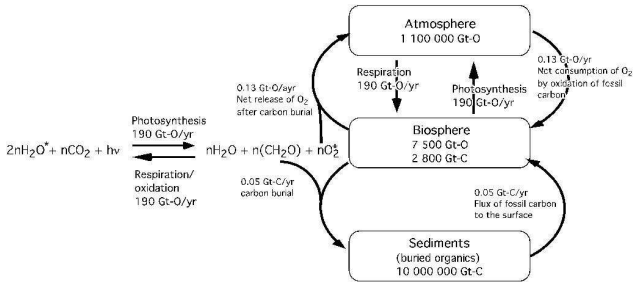}
\caption{Oxygen Cycle on Earth}
\end{figure}
%__________________________________________________________________

\section{Cloud features}
Clouds are an important component of exoplanet spectra because in the visible/NIR range, their reflection is high and relatively 
flat with wavelength, while in the infrared they lower the emitted flux and hide the lower convective region of the atmosphere 
that produce most of the spectral features (due to the strong gradient of temperature associated with the convection). Clouds 
hide the atmospheric molecular species below them, weakening the spectral lines in both the thermal infrared and visible. In the 
thermal infrared, clouds emit at temperatures that are generally colder than the surface, while in the visible the clouds 
themselves have different spectrally-dependent albedos that further influence the overall shape of the spectrum. Fig.4 shows 
the visible and thermal infrared spectral emission of the Earth for 3 cloud conditions (a) and the model to data comparison (b) 
\cite{Kaltenegger2006}. 

%                                                One column figure
%----------------------------------------------------------- Fig2
\begin{figure}
\centering
\includegraphics[width=8.4cm]{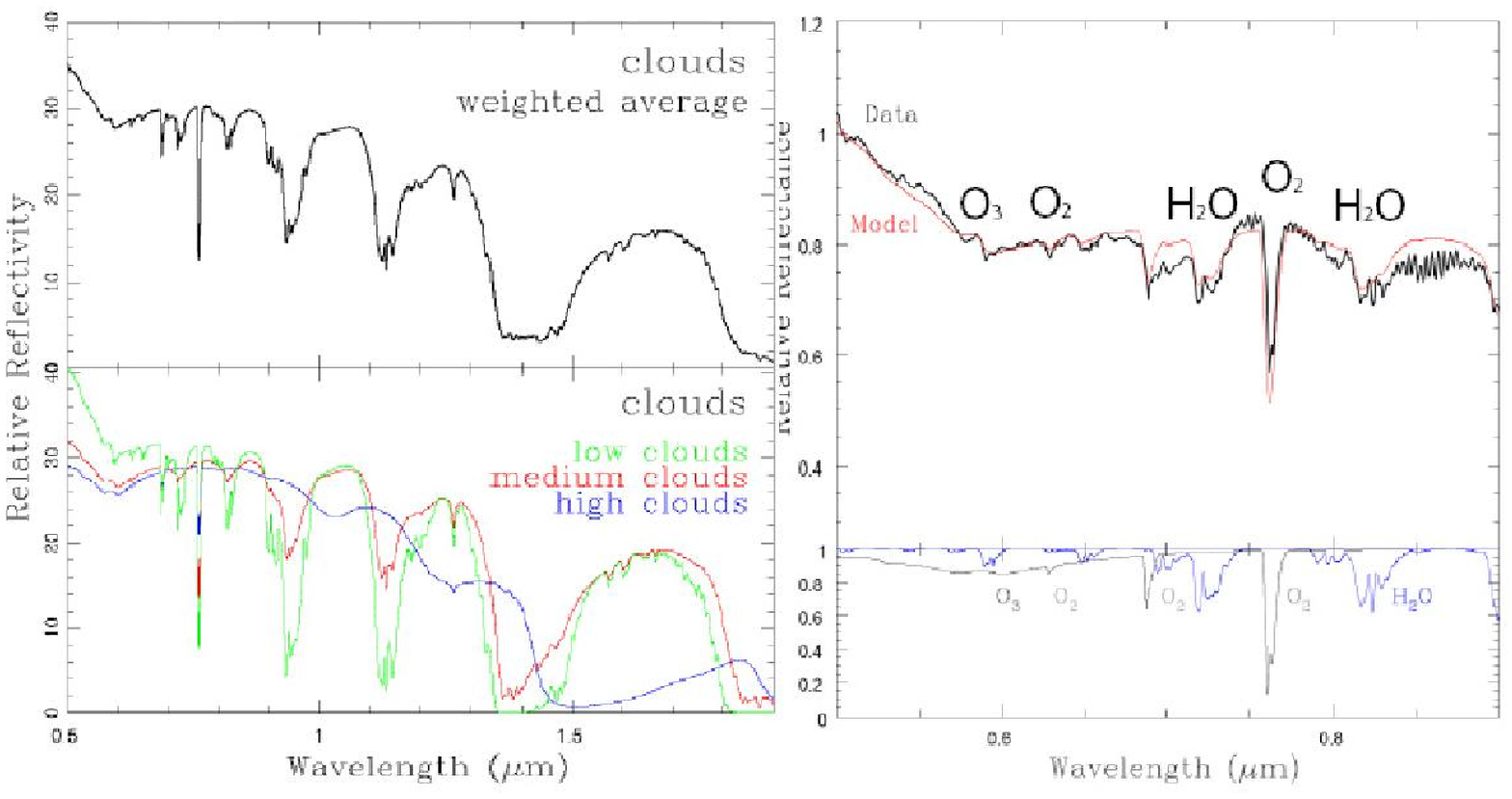}
\includegraphics[width=8.4cm]{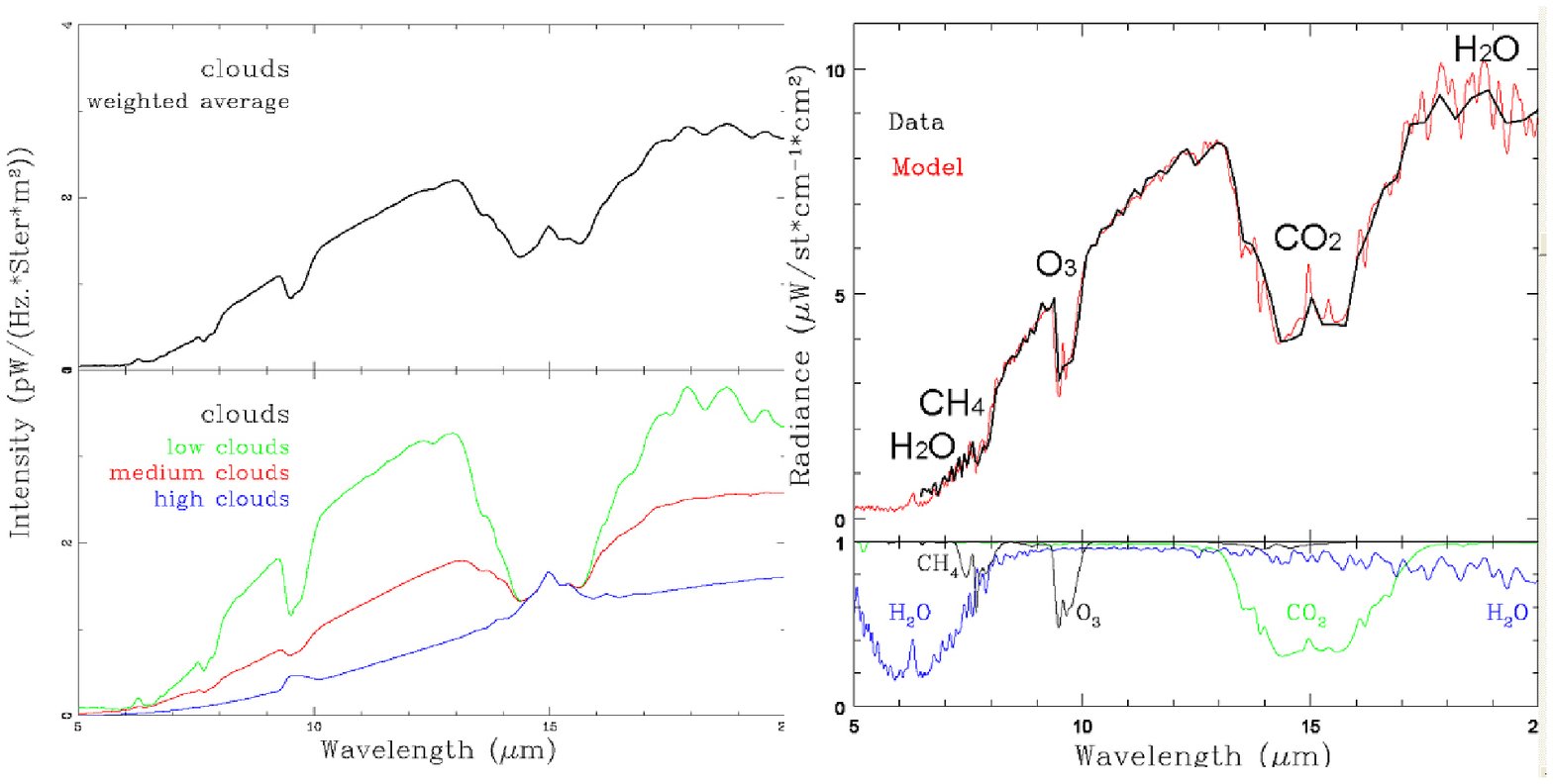}
\caption{(left) Spectra of present-day Earth with (green) 100\% cumulus cloud coverage at 1 km and (red) 100\% cumulus cloud 
coverage at 6 km and (blue) 100\% cirrus cloud coverage at 12 km, and, spectra of a mixture of clouds resembling the 
present Earth in the visible (upper panel) and in the thermal infrared (lower pannel).
Model Data comparison (b) data in black \cite{Woolf2002},\cite{Christensen1997} model in red \cite{Kaltenegger2006}.}
\end{figure}
%
%__________________________________________________________________

\section{Biomarkers and their evolution over geological times on Earth}
The spectrum of the Earth has not been static throughout the past 4.5 Ga (Ga = $10^9$ years ago). This is due to the variations in the molecular 
abundances, the temperature structure, and the surface morphology over time. Kasting and Catling \cite{Kasting2003} and Kasting \cite{Kasting2004} 
established a scenario for the Earth's atmosphere evolution. At about 2.3 Ga oxygen and ozone became abundant, 
affecting the atmospheric absorption component of the spectrum. At about 2 Ga, a green phytoplankton signal developed in the 
oceans and at about 0.44 Ga, an extensive land plant cover followed, generating the red chlorophyll edge in the reflection 
spectrum. The composition of the surface (especially in the visible), the atmospheric composition, and temperature-pressure 
profile can all have a significant influence on the detectabilty of a signal. Note that we assume that the cloud cover over 
these epochs is the same as the cloud cover today, however changes in this distribution could significantly change the overall 
spectra in both wavelength regions. 

Fig.5 (a) show theoretical visible and mid-infrared spectra of the Earth at six epochs during its geological evolution and (b) the 
required resolution to match the main spectral features and biomarkers of atmospheric compounds over geological time 
for Earth \cite{Kaltenegger2006}. The epochs are chosen to represent major developmental stages of the Earth, and life on 
Earth. Earth's atmosphere has experienced dramatic evolution over 4.5 billion years, and other planets may exhibit similar or 
greater evolution, at similar or different rates. The climate model to create a schematic atmospheric model of our Earth over 
geological timescales is based on a combination of results from work by Kasting and Catling \cite{Kasting2003}, Kasting \cite{Kasting2004}, 
Pavlov et al. \cite{Pavlov2003}, Segura et al. \cite{Segura2003} and Traub and Jucks \cite{Traub2002}. The model atmosphere evolves from a $CO_2$ rich atmosphere (3.9 Ga = epoch 0) to a $CO_2$/$CH_4$-rich atmosphere (epoch 3) to a present- day atmosphere (epoch 5 = present-day Earth). 
It shows epochs that reflect significant changes in the chemical composition of the atmosphere. 
The oxygen and ozone absorption features could have been used to indicate the presence of biological activity on Earth anytime 
during the past 50\% of the age of the solar system. The dark lines show a resolution of 70 and 25, as proposed for TPF-C in the visible
and Darwin/TPF-I in the IR respectively. 

Different signatures in the atmosphere are clearly visible over earth's evolution and observable with low resolution.
Only in the last epoch (5), vegetation modifies the reflected spectrum of earth by introducing a sharp increase of the 
reflectivity between 700-750 nm, the red edge. Its detection is extremely challenging in a spatially unresolved global
planetary spectrum, integrated over a long exposure 
\cite{Kaltenegger2006},\cite{Paillet2006}, \cite{Arnold2002}, \cite{Montanes2006}, \cite{Montanes2007},\cite{Seager2002}.
The spectral resolution required for optimal detection of habitability and biosignatures
has to match those features on our own planet for the dataset we have over its evolution.

%                                                One column figure
%----------------------------------------------------------- Fig3
\begin{figure}
\centering
\includegraphics[width=8.4cm]{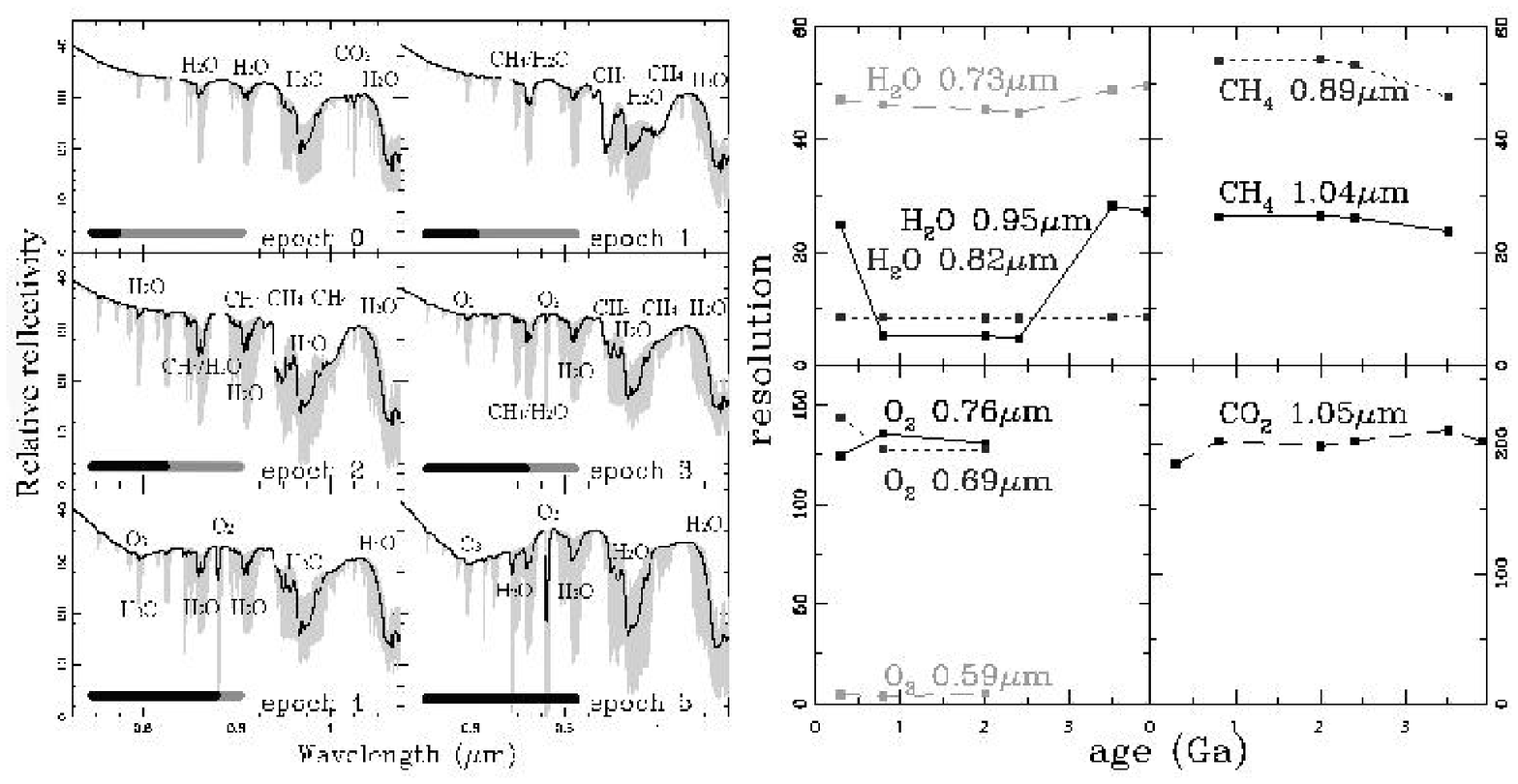}
\includegraphics[width=8.4cm]{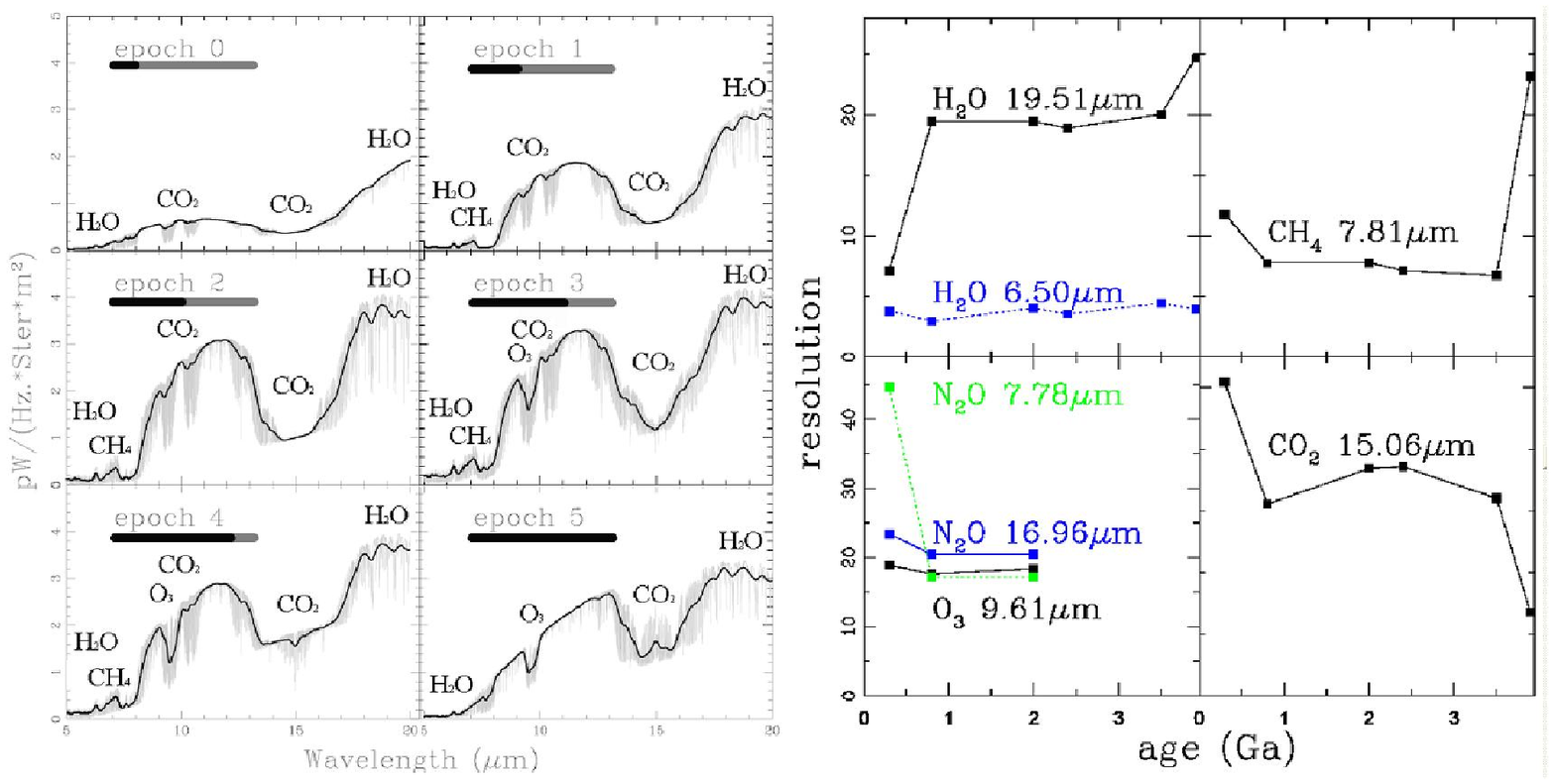}
\caption{The visible (upper pannel) and mid IR (lower pannel) spectral features on an Earth-like planet change considerably over its 
evolution from a $CO_2$ rich (epoch 0) to a $CO_2$/$CH_4$-rich atmosphere (epoch 3) to a present-day atmosphere (epoch 5) 
(black lines shows spectral resolution of 70 for the visile and 25 for the mid-IR). (b) Required resolution to match the main 
spectral features and biomarkers of atmospheric compounds over geological time for Earth \cite{Kaltenegger2006}}.
\end{figure}
%
%__________________________________________________________________

\section{Planets around different stars}
For an Earth-like planets in the HZ around a given star, the thermal flux will to first order be constant for a given planetary 
size, while the reflected stellar flux will scale with the brightness of the star. The suppression of the primary's thermal 
emission will, on the other hand, be progressively easier for later and later spectral types. The contrast ratio is a factor of 
about 2 more for FV stars, while it is a factor of 3 less for KV stars and about 11 less for MV stars in the N band compared 
to the sun-Earth \cite{Kaltenegger2007}. Surprisingly enough, it may thus be easier for the IR interferometer concept to detect a habitable 
Earth around an M-Dwarf than around something more akin to our own Sun. This is true for interferometric systems like Darwin and 
TPF-I that can be adapted to each individual target system, since the HZ moves closer and closer to the star for later and later 
stellar types. The baseline of the interferometers have to increase to resolve M star planetary-systems at larger distances, a 
constraint that is taken into account for the M target systems in the target star catalogue \cite{Kaltenegger2007}.

Using a numerical code that simulates the photochemistry of a wide range of planetary atmospheres, Selsis \cite{Selsis2000} and Segura et al \cite{Segura2003}, \cite{Segura2005} 
have simulated a replica of our planet orbiting different types of star: F-type star (more massive and hotter than the Sun) and 
a K-type star (smaller and cooler than the Sun). The models assume the same background composition of the atmosphere as well as the 
strength of biogenic sources. The orbital distance was chosen in order to give the planet the Earth's effective temperature (by 
receiving the same energetic flux): 1.8 and 0.5 AU respectively for the F- and the K-type star. Scaling the energetic flux allows 
us to consider habitable planets irradiated by a non-solar spectrum: the contribution of the UV range (150-400 nm, the most 
important for the photochemistry) is higher for the F-type star and lower for the K-type star (this is no longer true in the EUV 
range, below 150 nm, where low-mass stars like K-type stars are very active).

The results of modeling, illustrated in Fig.6, show the changes in detectability and shape of spectral features due to ozone, 
carbon dioxide, and methane for the 'same' planet around stars of different spectral type. These changes 
are due to an interplay between the star's spectrum, the photochemistry of ozone, and coupled changes in the thermal structure 
of the planet's atmosphere. These models were run for host stars of F, G, K \cite{Segura2003}, \cite{Selsis2000} and M spectral 
type \cite{Segura2005} and show that in high resolution the detectable features around e.g, a K star are deeper than the 
features around an F host star.

%                                                One column figure
%----------------------------------------------------------- Fig4
\begin{figure}
\centering
\includegraphics[width=8.cm]{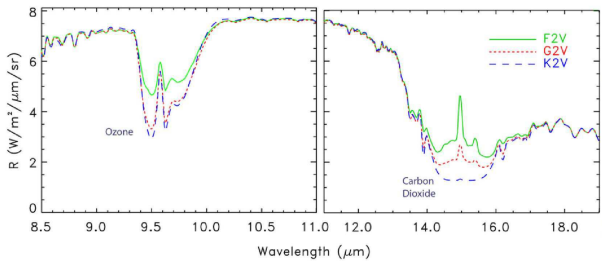}
\caption{These spectra show the appearance of Earth-like planets orbiting within the habitable zones of stars of F2V, G2V and 
K2V stellar type. In each case, a weakly-coupled radiative/photochemical atmospheric model was used to determine the equilibrium 
atmospheric composition and vertical structure for a planet with Earth's modern atmospheric composition, and radiatively forced 
by the UV to far-infrared spectrum of a star of each spectral type \cite{Segura2005}}.
\end{figure}
%
%__________________________________________________________________

\section{Abiotic sources}
We need to address the abiotic sources of biomarkers, so that we can identify when it might constitute a 'false positive' for life. 
$CH_4$ is an abundant constituent of the cold planetary atmospheres in the outer solar system. On Earth, it is produced abiotically 
in hydrothermal systems where $H_2$ (produced from the oxidation of Fe by water) reacts with $CO_2$ in a certain range of pressures 
and temperatures. In the absence of atmospheric oxygen, abiotic methane could build up to detctable levels. Therefore, the sole 
detection of $CH_4$ cannot be attributed unambiguously to life.

$O_2$ also has abiotic sources, the first one is the photolysis of $CO_2$, followed by recombination of O atoms to form $O_2$ $(O + O + M \rightarrow O_2 + M)$, a second one is the photolysis of $H_2O$ followed by escape of hydrogen to space. The first source is a steady state maintained by 
the stellar UV radiation, but with a constant elemental composition of the atmosphere while the second one is a net source of 
oxygen. In order to reach detectable levels of $O_2$ (in the reflected spectrum), the photolysis of $CO_2$ has to occur in the 
absence of outgassing of reduced species and in the absence of liquid water, because of the rainout of oxidized species. Normally, 
the detection of the water vapor bands simultaneously with the $O_2$ band can rule out this abiotic mechanism (Segura et al. in preparation). 
In the infrared, this process cannot produce a detectable $O_3$ feature \cite{Selsis2002}.
The loss of hydrogen to space can result in massive oxygen leftovers: 240 bars of oxygen could build up after the loss of the 
hydrogen contained in the Earth ocean. However, the case of Venus tells us that such oxygen leftover has a limited lifetime in 
the atmosphere (because of the oxidation of the crust and the loss of oxygen to space): we do not find $O_2$ in the Venusian 
atmosphere despite the massive loss of water experienced in the early history of the planet. 
Also, such evaporation-induced build up of $O_2$ should occur only close to the star (see Fig.2) and affect small planets with low gravity more 
dramatically. For small planets ( $<$ 0.5 $M_{Earth}$) close to inner edge of the habitable zone ($<$ 0.93 AU from the present 
Sun), there is a risk of abiotic oxygen detection, but this risk becomes negligible for bigger planets further away from their star.

\section{Biomarkers detection set in context}
The search for signs of life implies that one needs to gather as much information as possible in order to understand how the 
observed atmosphere physically and chemically works. Knowledge of the temperature and planetary radius is crucial for the 
general understanding of the physical and chemical processes occurring on the planet (tectonics, hydrogen loss to space). In 
theory, spectroscopy can provide detailed information on the thermal profile of a planetary atmosphere. This however 
requires a spectral resolution and a sensitivity that are well beyond the performance of a first generation spacecraft like 
Darwin/TPF-I and TPF-C. Thus we will concentrate on the initially available observations here like flux variations of a planet 
throughout its orbit.Smaller coronagraphs or interferometers can probe similar characteristics for extrasolar giant planets. 

\subsection{Temperature and Radius of the Planets}
One can calculate the stellar energy of the parent star that is received at the planet's measured orbital distance. This only gives very 
little information on the surface temperature of the planet, which depends on its albedo and the efficiency of the greenhouse 
effect. However, with a low resolution spectrum of the thermal emission, the mean effective temperature and the radius of the 
planet can be obtained, in first approximation, by fitting the envelope of the thermal emission by a Planck function. The ability 
to associate, by doing such fit, a surface temperature to the spectrum relies on the existence and identification of spectral 
windows probing the surface or the same atmospheric levels. For an Earth-like planet there are some atmospheric windows that can 
be used in most of the cases, especially between 8 and 11 $\mu$m as seen in Fig.1 and 3. This window would however become opaque at 
high $H_2O$ partial pressure (e.g. inner part of the HZ where a lot of water is vaporized) and at high $CO_2$ pressure (e.g. a very 
young Earth (see Fig.3) or outer part of the HZ). A much better estimate of the radius and of the temperature can be obtained 
by comparing the spectrum, including its features, with a grid of synthetic spectra \cite{Paillet2006}, (Kaltenegger,Selsis, in prep).

\section{Orbital flux variations}
The orbital flux variation in the IR, measured in the detection phases, can provide some precious information about the planet. The 
thermal light curve (i.e. the integrated infrared emission measured at different position on the orbit) exhibits variations due 
to the phase (whether the observer sees mainly the day side or the night side) and to the season (if the planet has a non zero 
obliquity). Important phase-related variations are due to a high day/night temperature contrast and imply a tenuous or no 
atmosphere and the absence of a stable liquid ocean. Therefore, habitable planets can be distinguished from airless or 
Mars-like planets by the amplitude of the observed variations \cite{Selsis2003}, \cite{Gaidos2004}, see Fig.5. 
Note that also Venus-like atmosphere would exhibit extremely low amplitudes and can only be distinguished by spectroscopy from 
habitable planets. 

%                                                One column figure
%----------------------------------------------------------- Fig5
\begin{figure}
\centering
\includegraphics[width=8.cm]{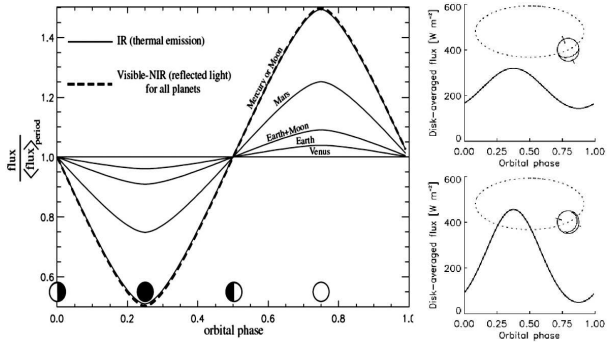}
\caption{Orbital light curve for black body planets in a circular orbit with and without an atmosphere in the thermal infrared \cite{Selsis2003} (left) and the influence of different rotation axis on the shape of the lightcurve \cite{Gaidos2004} (right).} 
\label{Fig5}
\end{figure}
%
%__________________________________________________________________

The mean value of $T_b$ estimated over an orbit can be used to estimate the albedo of the planet, $A$, through the balance between the incoming stellar radiation and the outgoing IR emission.

\begin{equation}
F_{star} (1 - A) = 4  <T_b>^4_{orbit}
\end{equation}

In the visible ranges, the reflected flux allows us to measure the product $A × R^2$, where $R$ is the planetary radius (a small 
but reflecting planet could be as bright as a big but dark planet). The first generation of optical instruments will be very 
far from the angular resolution required to directly measure an exoplanet radius. Presently, such a measurement can only be 
performed when the planet transits in front of its parent star, by an accurate photometric technique. If the same target is 
observed in both visible (TPF-C) and IR (Darwin/TPF-I) ranges, the albedo can be obtained in the visible once the radius is 
inferred from the IR spectrum. The measurement of the albedo in the visible can then be compared to the albedo estimated from 
the IR lightcurve and the results can be iterated. The possibility to obtain spectral information from both wavelengths will 
allow a more detailed characterization of individual planets and it also allows exploring a wide domain of planet diversity. 
Observations in both complementary wavelength bands can confirm the presence of atmospheric compounds. Some important species 
like $CO_2$ and $N_2O$ appear only in the IR range, while only the reflected spectrum can give information on the nature of 
the surface. Information on the cloud and surface characteristics (ocean, ice, rocks) can be obtained once the absolute level 
of the albedo is known, which requires the knowledge of the radius. At extremely high temporal resolution
and signal-to-noise ratio additional modulations of the amplitude variations of the visual reflected
light over a daily rotation of a planet could indicate surface features \cite{Ford2001}. As shown in the same
paper, clouds destroy that relation, because of their own individual rotation pattern. 

The outgoing short-wave and long-wave radiation, combined with their variations along the orbit, can
constrain albedo, greenhouse gases and would allow to explore climate systems at work on the observed 
worlds. 

\section{Summary}
Any information we collect on habitability, is only important in a context that allows us to interpret, what we find. The search 
for signs of life implies to gather as much information as possible in order to understand how the observed atmosphere physically 
and chemically works.  Knowledge of the temperature and planetary radius is crucial for the general understanding of the 
physical and chemical processes occurring on the planet. These parameters as well as an indication of habitability can be 
determined with low resolution spectroscopy and low photon flux, as assumed for first generation space missions. 

Our search for signs of life is based on the assumption that extraterrestrial life shares fundamental characteristics with life 
on Earth, in that it requires liquid water as a solvent and has a carbon-based chemistry. We adopt this conservative approach 
to rule out false positives, completely aware that we will miss to label some planets as habitable, where life has developed 
based on other chemistry or where the environment does not lead to a buildup of oxygen in the atmosphere. An amazing feature of 
future space based missions like Darwin and TPF is that they will have the capability, for the first time, to do comparative planetology 
on a wide variety of planets, whose atmosphere and climatology is far outside our current understanding, as well as find 
planets similar to our own and probe them for habitable conditions.

\acknowledgments
This work was sponsored by NASA grant NAG5-13045.

%\newpage
%A list of references on a section level follows. 
%\setlength{\bibindent}{4mm} % indentation for two digits

\end{document}